\documentclass[conference]{IEEEtran}
\IEEEoverridecommandlockouts
\usepackage{cite}
\usepackage{amsmath,amssymb,amsfonts}
\usepackage{algorithmic}
\usepackage{graphicx}
\usepackage{textcomp}
\usepackage{xcolor}
\usepackage{url}
\usepackage[nolist]{acronym}% consistently use acronyms
\def\BibTeX{{\rm B\kern-.05em{\sc i\kern-.025em b}\kern-.08em
    T\kern-.1667em\lower.7ex\hbox{E}\kern-.125emX}}
\begin{document}

\newacro{hmd}[HMD]{Head Mounted Display}
\newacro{VR}[VR]{Virtual Reality}
\newacro{fov}[FOV]{Field-of-View}
\newacro{ar}[AR]{Augmented Reality}
\newacro{xr}[XR]{Extended Reality}
\newacro{mr}[MR]{Mixed Reality}
\newacro{uml}[UML]{Unified Modeling Language}
\newacro{omg}[OMG]{Object Management Group}
\newacro{bpmn}[BPMN]{Business Process Model Notation}
\newacro{iot}[IoT]{Internet of Things}
\newacro{cave}[CAVE]{Cave Automatic Virtual Environment}
\newacro{fsm}[FSM]{Finite State Machine}
\newacro{soc}[SoC]{System on a Chip}
\newacro{pun}[PUN2]{Photon Unity Networking 2}
\newacro{hlapi}[HLAPI]{High Level Application Programming Interface}
\newacro{llapi}[LLAPI]{Low Level Application Programming Interface}
\newacro{sdk}[SDK]{Software Development Kit}
\newacro{gui}[GUI]{Graphical User Interface}
\newacro{rpc}[RPC]{Remote Procedure Call}
\newacro{sus}[SUS]{System Usability Scale}
\newacro{apis}[APIs]{Application Programming Interfaces}

\title{Collaborative Software Modeling in Virtual Reality
}

%\author{Anonymous}

\author{\IEEEauthorblockN{Enes Yigitbas, Simon Gorissen, Nils Weidmann, Gregor Engels}
\IEEEauthorblockA{\textit{Paderborn University}, Germany \\
firstname.lastname@upb.de}}

\maketitle

\begin{abstract}
Modeling is a key activity in conceptual design and system design. Through collaborative modeling, end-users, stakeholders, experts, and entrepreneurs are able to create a shared understanding of a system representation. While the Unified Modeling Language (UML) is one of the major conceptual modeling languages in object-oriented software engineering, more and more concerns arise from the modeling quality of UML and its tool-support. Among them, the limitation of the two-dimensional presentation of its notations and lack of natural collaborative modeling tools are reported to be significant. In this paper, we explore the potential of using Virtual Reality (VR) technology for collaborative UML software design by comparing it with classical collaborative software design using conventional devices (Desktop PC / Laptop). For this purpose, we have developed a VR modeling environment that offers a natural collaborative modeling experience for UML Class Diagrams. Based on a user study with 24 participants, we have compared collaborative VR modeling with conventional modeling with regard to efficiency, effectiveness, and user satisfaction. Results show that the use of VR has some disadvantages concerning efficiency and effectiveness, but the user’s fun, the feeling of being in the same room with a remote collaborator, and the naturalness of collaboration were increased.
\end{abstract}


\section{Introduction}

In modern software development collaboration between developers is one of the driving factors that determines the quality and speed at which the projects can be realized. One central artifact of communication and discussion in software engineering are models \cite{Whitehead2007}. The \ac{uml} with its associated diagrams is one of the most well known general-purpose modeling languages in software engineering and is considered by many as "lingua franca" for software engineers \cite{Petre2013}. However, researchers and software designers
have realized the insufficiency of \ac{uml} in its expressiveness since it is restricted to a two-dimensional plane. These insufficiencies include a lack of dynamic expression and interaction ability between groups of remote designers ~\cite{10.5555/1767297.1767349}, \cite{Maletic01visualizingsoftware} as well as the complexity for large models \cite{EricksonS03}. Furthermore, the authors in \cite{DBLP:conf/models/BadreddinKFML18} argue that the dissatisfaction of developers with \ac{uml} tools is one of the reasons it has not been adopted more universally, exemplifying the need for improving the tool support. Since the COVID-19 pandemic has spread around the world, this need for good \ac{uml} tool support has only increased. Many educational institutions, like schools and universities, around the world have been forced to switch to online education settings to support social distancing. Likewise, millions of workplaces wherever possible were transitioned to home office. To enable collaboration for software engineers in such situations, tools are needed that offer support for creating and discussing models from remote locations. While classical modeling applications, like \textit{Lucidchart} \cite{LucidCharts} or \textit{GenMyModel} \cite{GenMyModel}, support remote collaboration, they do not overcome the mentioned issues with regard to visualization and collaboration as they are mostly relying on a 2D UML notation and do not support a natural way of collaboration comparable to editing a model on a whiteboard while situated together in one room.

Virtual Reality technology, on the other hand, is becoming increasingly sophisticated and cost-effective and can be applied to many areas such
as training~\cite{DBLP:conf/vrst/YigitbasJSE20}, robotics \cite{DBLP:conf/seams/YigitbasKJE21}, education~\cite{DBLP:conf/mc/YigitbasTE20}, healthcare~\cite{DBLP:conf/mc/YigitbasHE19}, and even Information Systems (IS) research, to simulate a real environment or represent complicated scenarios. Modern \ac{hmd} \ac{VR} devices have several technological capabilities that are not present on conventional devices (Desktop PC / Laptop): (1) Stereoscopic 3D Images, (2) Six Degrees of Freedom, (3) Hand Presence Support. and (4) 3D Spatial Voice Chat. Since \ac{hmd} devices show a slightly different image to each eye, they can invoke the perception of a truly 3D virtual world that the user inhabits. These \textsc{Stereoscopic 3D Images} are more in line with the visual experience of the real world compared to images on a 2D screen, since people also perceive the real world in 3D, not in 2D. In \ac{VR}, users can also look around in the virtual world by simply moving their heads, like they would in a real environment. In addition to this rotational movement, modern \ac{VR} devices can also track the positional movement of a user, independent of the direction the user moves in. This means, the user has \textsc{Six Degrees of Freedom} (three rotational and three positional axes) in her movement and the \ac{VR} system can adapt the view and position of the user in the virtual world accordingly. Additionally, typical \ac{VR} controllers are used single-handedly with each controller representing one hand in the virtual world. This \textsc{Hand Presence Support} allows a user to make natural gestures like grabbing an object and pointing at things to interact with the virtual world. With the means of \textsc{3D Spatial Voice Chat}, it is possible to make users feel like the voices of their collaborators come with the volume level (distance-based) and from the direction that their virtual representations (i.e. Avatars) are in. Overall, the fast development in VR concerning prototyping \cite{DBLP:conf/hcse/JovanovikjY0E20} and engineering of VR applications \cite{DBLP:journals/corr/abs-2107-00377} as well as the mentioned technological advances allow us to extend the research field to improve the quality of UML and collaborative modeling. Thus, the main goal of this paper is to explore the potential of using Virtual Reality (VR) technology for collaborative UML software design by comparing it with classical collaborative software design using conventional devices (Desktop PC / Laptop). For this purpose, we have developed a VR modeling environment, called \textit{VmodlR}, that offers a natural collaborative modeling experience for UML Class Diagrams. Based on a user study with 24 participants, we have compared collaborative VR modeling with conventional modeling with regard to efficiency, effectiveness, and user satisfaction.

The rest of the paper is structured as follows. In Section 2, we present and discuss the related work. In Section 3, we describe the conceptual solution of our VR-based collaborative modeling environment \textit{VmodlR}. In Section 4, we show the details of the implementation of \textit{VmodlR}. In Section 5, we present and discuss the main results of the usability evaluation. In Section 6, we conclude the paper and give an outlook for future work.

\section{Related Work}

Model-based and model-driven development methods have been discussed in the past for various application domains such as intelligent user interfaces \cite{DBLP:conf/ucami/YigitbasGSE17, DBLP:conf/eics/Yigitbas0E17, DBLP:journals/sosym/YigitbasJBSE20}, usability engineering \cite{DBLP:conf/hcse/YigitbasAJK0E18, DBLP:journals/pacmhci/YigitbasHRASE19, DBLP:conf/eics/YigitbasM015}, digital twins \cite{DBLP:conf/hci/JosifovskaYE19, DBLP:conf/icse/JosifovskaYE19} or business model development \cite{DBLP:conf/caise/GottschalkYNE21, DBLP:conf/bmsd/GottschalkYE20}. In the following, we draw on prior research into \textit{Collaborative Modeling}, \textit{3D Modeling}, and \textit{Immersive Modeling}.

\subsection{\textit{Collaborative Modeling}}

In previous work, the topic of collaborative software modeling was already researched from different perspectives \cite{DBLP:journals/ijspm/RengerKV08}. In \cite{Boul2002}, the authors present a distributed \ac{uml} editor that aims to transfer collaborative discussion and editing of \ac{uml} models from regular meeting rooms to remote settings using regular computers. In addition, a collaborative learning environment for UML modeling is presented in \cite{DBLP:journals/ile/ChenPP06}. Furthermore, \cite{DBLP:conf/vissoft/FerencPV17} introduces an approach to a real-time synchronous collaborative modeling of software systems using 3D UML.
Besides these approaches from research, there are also many commercial tools supporting collaborative software modeling. \textit{Lucidchart} is a web-based commercial tool for collaboratively creating diagrams \cite{LucidCharts}. It includes support for many diagram types and modeling languages like \ac{uml}, \ac{bpmn}, Enterprise Relationship models, and more. The tool shows a 2D canvas where standard 2D diagrams can be created. Since Lucidchart is web-based, it can be used by any device that runs a modern browser, although as with many web-based contents the experience can be assumed to be best suited to the use on desktop and laptop computers and is not specifically adopted to the possibilities of \ac{VR} devices. Released by Axellience in 2014, GenMyModel \cite{GenMyModel} is a web-based tool similar to Lucidchart. GenMyModel supports a subset of the most commonly used UML diagrams, i.e., Class, Use Case, Component, Object, State, Deployment, Activity, and Sequence Diagrams. Axellience describes the mode of collaboration as similar to how Google Drive or Microsoft Office online collaboration works. So users can see where their co-workers are editing the model but do not have an integrated voice chat to discuss those changes. This has to be done via an external tool. The system aims to support the language's visual representations according to their official definitions and thus only presents 2D models to the users. 

While the above-described approaches mostly support collaborative modeling based on standard 2D \ac{uml} diagrams, they do not support immersive VR.

\subsection{3D Modeling}

The authors in \cite{10.1145/345513.345358} presented a conceptual system that visualizes \ac{uml} Class and Sequence Diagrams in 3D. The Sequence Diagrams are displayed in the context of the Classes they belong to and animated to emphasize their connection. Furthermore, the authors in \cite{Zhang05} have also discussed a conceptual approach to extending 2D \ac{uml} Diagrams to the third dimension. They introduce some modeling quality attributes and explain how \ac{VR} can improve the \ac{uml} modeling quality according to these attributes. For example, they argue that the model's "understandability" can be improved through \ac{VR}'s "Immersion" and "Stereopsis" (Stereoscopic 3D presentation) features. Subsequently, they provide some examples of a Class Diagram and a Sequence Diagram that are visualized in 3D and demonstrate their advantages in comparison to a 2D representation. In \cite{Rod2016}, the authors proposed VisAr3D, a 3D visualization tool for \ac{uml} models to make large models easier to comprehend, especially by inexperienced modelers like students. The system automatically converts a 2D \ac{uml} model from an .xmi file into its 3D representation. The virtual environment the 3D \ac{uml} models are placed in can be viewed through a web-based app and does not support \ac{hmd}s. Their prototype is not a model editor, however, it only visualizes \ac{uml} models in 3D. This approach was evaluated based on a user study assessing the effect of the 3D compared to a 2D representation. The main results show that the 3D representation aided the understandability of large models, "increased students' interest" and supported teaching purposes \cite{Rod2016}. In \cite{Leyer2019}, the authors described an implementation that visualized a process model in a 3D virtual world on a desktop computer. It is a training system intended to aid a single user in understanding Business Process Models through 3D visualization. They evaluate how the 3D representation assists the user in learning the modeled process compared to a standard 2D model representation. They conclude that the 3D representation provides a noticeable learning benefit. It is not a modeling tool, however, only a visualization and training tool. Further examples for approaches which make use of a 3D representation of UML models can be found in \cite{DBLP:conf/pppj/CaseyE03, DBLP:conf/softvis/PilgrimD08, DBLP:journals/ivs/PilgrimDM09, DBLP:conf/aswec/McIntoshH10}. In summary, a modeling approach that combines and integrates the aspects of 3D modeling, collaboration, and VR in one solution is not fully covered and yet existing.

\subsection{Immersive Modeling in AR and VR}

In the following, AR- and VR-based approaches for modeling purposes will be briefly described and discussed. Although VR is our main focus, we included AR approaches to cover immersive modeling approaches on the whole. 

\ac{ar} is closely related to \ac{VR} with the main difference being that \ac{VR} immerses a user in a completely virtual world while \ac{ar} does not isolate the user from the real world by displaying virtual objects in the real environment. In general, AR has been already applied for different aspects such as robot programming \cite{DBLP:journals/corr/abs-2106-07944}, product configuration (e.g., \cite{DBLP:conf/hcse/GottschalkYSE20}, \cite{DBLP:conf/hcse/GottschalkYSE20a}), planning and measurements \cite{DBLP:conf/eics/EnesScaffolding} or for realizing smart interfaces (e.g., \cite{DBLP:conf/eics/KringsYJ0E20}, \cite{DBLP:conf/interact/YigitbasJ0E19}). To be more specific, example approaches that apply AR for software modeling are as follows. The authors in \cite{Mikkelsen2017} have presented a framework that is supposed to allow editing and viewing \ac{uml} models in a 3D space through the Microsoft Hololens AR Glasses. Their approach is to overlay 3D model elements over real scenery so the user can move around and inside the model while not being shut off from the real world. In this way, their prototype only displays static objects that are meant to represent \ac{uml} models. Similarly, in \cite{Reuter2019}, the authors proposed a system that aims to make learning \ac{uml} more accessible by displaying it in 3D as overlays to the natural environment using the Microsoft Hololense. The system supports creating and editing \ac{uml} Class Diagrams but is only intended for single users and does not support collaboration. Furthermore, in \cite{seiger2021holoflows} the authors introduce the concept of "HoloFlows" to support the modelling of processes for the internet of things in mixed reality. A similar solution is introduced in \cite{DBLP:conf/icse/BrunschwigCGL21} where the authors present an approach for supporting domain-specific modelling environments based on AR. The main drawback of AR-based solutions for modeling is the small field of view which narrows down the possibilities for modeling support. Therefore, we have explored an alternative solution in VR.

Focusing on \ac{VR}-related approaches for software modeling, we can see that many previous works already have seen the potential in using \ac{VR} for improving modeling activities. In \cite{Malet2001}, for example, the authors have presented a system that analyses an object-orientated code base and visualizes its classes, attributes, and relations automatically in a 3D virtual environment that users can inspect inside a \ac{cave}. So it supports \ac{VR} only in a broader sense. The application is not networked and does not support editing the model, it is only a visualization tool. In \cite{Dengel2018}, the authors have proposed a system for \ac{hmd} \ac{VR} devices that can import a \ac{fsm} and visualize it in a game-like environment where players stand on islands representing states and can change islands via different boats representing the possible transitions. It is a single-player game environment meant for educational purposes and does not support UML modeling with actual \ac{uml} Diagram elements or remote collaboration. In \cite{Oberhauser2018}, the authors have presented a \ac{hmd} \ac{VR} system for visualizing process models in 3D. Therein, the models can be annotated but not edited. They evaluated the effectiveness, efficiency, and intuitiveness of the \ac{VR} visualization in comparison to (1) paper and (2) desktop tool based \ac{bpmn} and found that the effectiveness was equivalent between \ac{VR} and desktop but task completion in \ac{VR} was faster than with the \ac{bpmn} tool and the users found the \ac{VR} controls very intuitive.

% \begin{table}[htbp]
% \caption{Overview of the fulfilled requirements in the related work}
% \begin{center}
% \begin{tabular}{|c|c|c|c|c|c|}
% \hline
% \textbf{Related}&\multicolumn{5}{|c|}{\textbf{Requirements}} \\
% \cline{2-6} 
% \textbf{Work} & \textbf{\textit{R1}}&\textbf{\textit{R2}}&\textbf{\textit{R3}}&\textbf{\textit{R4}}&\textbf{\textit{R5}} \\
% \hline
% Unity\cite{UnityWebpage} &  \checkmark &  \checkmark & \checkmark & O & O \\
% cooperate immersive tool\cite{ImmersiveEUD-MultiUser} &  \checkmark &  \checkmark & \checkmark & X & X \\
% GREP\cite{EUDGames-GREP} &  \checkmark &  X & X & O & \checkmark  \\
% Hubs\cite{HubsWebpage} \& Spoke\cite{SpokeWebpage}&  \checkmark &  \checkmark & \checkmark & X & X\\
% Sumerian\cite{SumerianWebpage} &  \checkmark &  \checkmark & O & \checkmark & O \\
% \hline
% \multicolumn{6}{l}{\checkmark = fulfilled  O = fulfilled partly X = not fulfilled}\\
% \end{tabular}
% \label{overviewRelated}
% \end{center}
% \end{table}

\section{Solution Overview}

\begin{figure}[h]
\centerline{\includegraphics[scale=0.27]{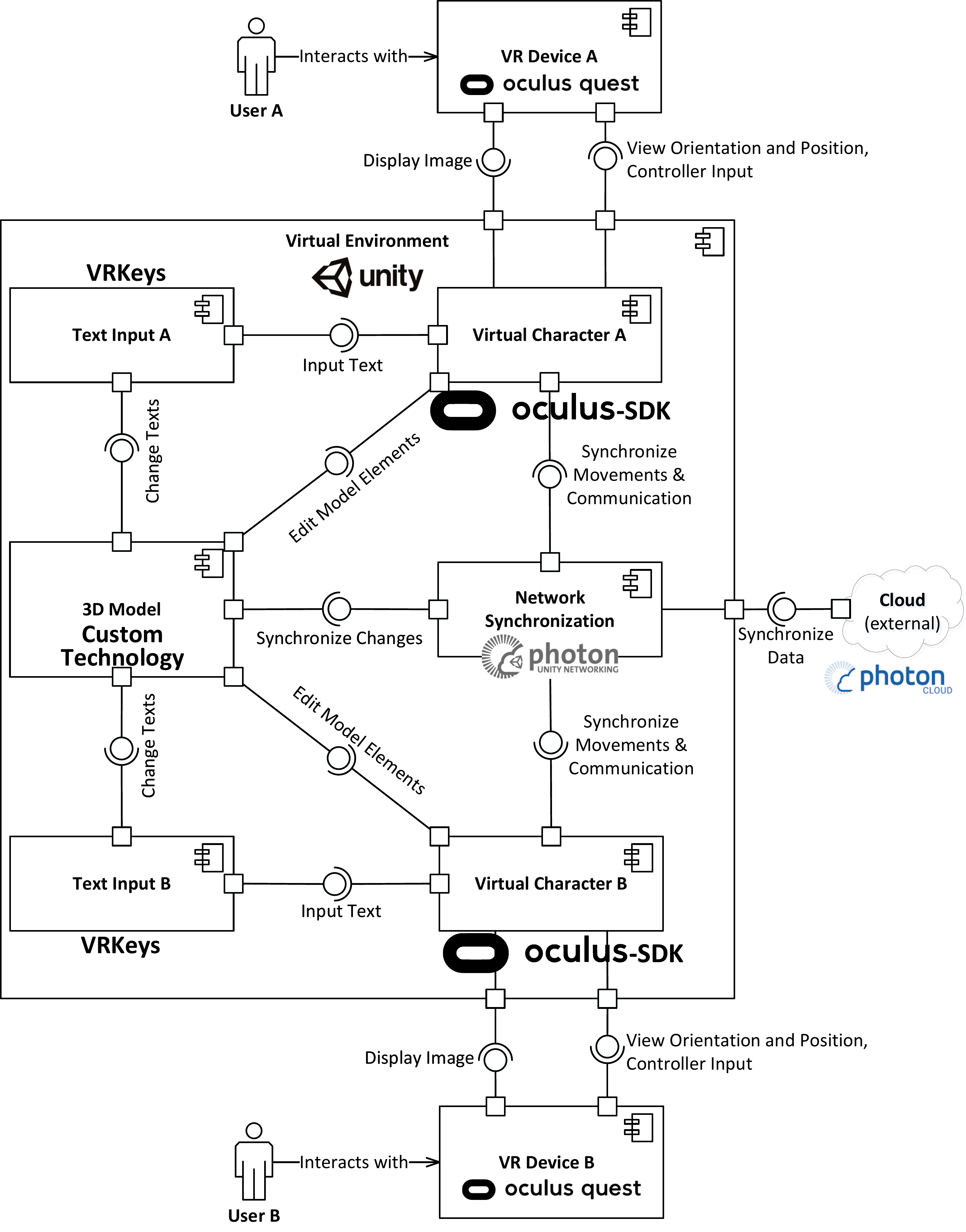}}
\caption{System overview of \textit{VmodlR}}
\label{fig:sysOver}
\end{figure}

The system overview of our VR-based collaborative modeling environment \textit{VmodlR} is shown in Fig. \ref{fig:sysOver}. The top half of this figure represents \textit{User A} and the \textit{Virtual Environment} she accesses through a \textit{VR Device}. The bottom half symmetrically shows a remote collaborator, \textit{User B}, and the \textit{VR Device} she uses to access the same shared \textit{Virtual Environment}. Through the \textit{\ac{VR} Device}, each \textit{User} is represented in the \textit{Virtual Environment} as a \textit{Virtual Character}. The \textit{View Orientation and Position} of the \textit{VR Device} together with the \textit{Controller Input} controls the \textit{Virtual Character} while the \textit{Virtual Environment} with its content is displayed from the \textit{Virtual Character}'s perspective inside the \textit{\ac{VR} Device}.

Through the \textit{Virtual Character}, each \textit{User} interacts with the elements inside the \textit{Virtual Environment}: The \textit{3D Model} can be edited by either directly editing model elements (e.g. creating, deleting, or moving them), or through the \textit{VR Text Input} component that is used to edit the text inside the \textit{3D Model}, for example, the Class names. All those changes to the model are synchronized between users through the \textit{Network Synchronization} component. In the case of this solution, the \textit{3D Model} is a three-dimensional \ac{uml} Class Diagram, but theoretically, this could be adapted to any kind of conceptual model. Through their \textit{Virtual Characters}, \textit{User}s can also interact with each other via \textit{Network Synchronization} which transmits their voices and synchronizes their body and hand movements. This is visualized in Fig. \ref{fig:sysOver} as the \textit{Synchronize Movements \& Communication} interface between the \textit{Virtual Character}s and the \textit{Network Synchronization}. The \textit{Network Synchronization} component then ensures that the \textit{Virtual Environment} and its content is synchronized between the \textit{User}s through the \textit{Cloud}. It offers several different services that can be used by other components to synchronize all necessary aspects of the \textit{Virtual Environment}. In the following, each of the main components \textit{Virtual Environment}, \textit{Network Synchronization}, \textit{Virtual Character}, \textit{3D Models}, and \textit{VR Text Input} will be described in more detail. 

\subsection{\textit{Virtual Environment}}

The \textit{Virtual Environment} consists of all virtual elements that are needed to provide a collaborative software modeling environment. Within this environment, each user can see and move around via their \textit{VR Device} in 3D. As a design decision for the \textit{Virtual Environment}, we opted to use an open space instead of a closed one to not introduce some unrealistic environment behavior or limit the user's ability to create large models. Since there is no open environment that could be considered natural for creating conceptual models, any space that offers a planar ground for the user to walk on could be chosen. We decided on a grass field under a blue sky because it does not limit the 3D space available to the user and depicts a pleasant real-world environment that users are familiar with.

\subsection{\textit{Network Synchronization}}

All elements inside the \textit{Virtual Environment} have to be synchronized through the \textit{Network Synchronization} component. We will briefly describe how this networking generally works in this conceptual solution. The networking has a server-based architecture where all users connect to a server and the server synchronizes instances of a \textit{Virtual Environment} between all users that are currently in that environment. It is important to understand that a networked environment is not a singular environment. Instead, on each user's \textit{\ac{VR} Device} (clients) a local version of the networked environment exists, so the user can look around in it and interact with it. All the changes the user can thereby make inside the \textit{Virtual Environment}, like moving their \textit{Virtual Character}, are communicated to the server so it can update its reference representation of the networked environment and forward the changes to the local environments of all other clients. These clients then apply the changes accordingly to their local copies of the environment. This way, the local environments on all clients are always kept in sync with the server's networked environment. Three components can be used to synchronize objects inside the \textit{Virtual Environment}: The \textit{Movement Synchronization Service}, the \textit{Event Synchronization Service}, and the \textit{Voice Synchronization Service}. The \textit{Movement}- and \textit{Event Synchronization Service}s can be more generally used. The difference between them is that the \textit{Movement Synchronization Service} is dedicated to rapidly and frequently changing information, that need to be synchronized many times per second. In the case of movement, this is needed to show the movement of an object that is moved by a remote collaborator fluently to the local user. The \textit{Movement Synchronization Service} also deals with tracking if a synchronized object that can be moved gets created or deleted. The \textit{Event Synchronization Service}, on the other hand, is supposed to synchronize arbitrary events that happen rather infrequently and therefore only have to be synchronized occasionally, once they occur, instead of the constant synchronization needed for movement. This provides flexibility where something that only occasionally changes (like the name of a Class for example) can be synchronized via the \textit{Event Synchronization Service} and otherwise does not consume network bandwidth while things that often change and have to be synchronized many times per second, like the movement of a Class or a \textit{Virtual Character}'s \textit{Virtual Hands}, can be synchronized fluently via the \textit{Movement Synchronization Service}. Finally, the \textit{Voice Synchronization Service} is dedicated to synchronizing the voices of users to enable voice chat inside our solution. 

\subsection{\textit{Virtual Character}}

In \ac{VR}, people can be represented by 3D characters through an avatar with a body, head, and hands. This way, a user can, for example, move around in the virtual environment and point with their finger in real life and the \textit{VR Controller} can reproduce this gesture on the \textit{Virtual Character}'s hands. Since this has the possibility to make discussions about models in \ac{VR} much more natural than possible on PCs, these features were also included in the design of our solution and the hands' gestures are synchronized across the network for each user, along with the positional audio of voices and the positions and orientations of \textit{Virtual Character}s. Our solution supports this form of natural movement, where the \textit{Virtual Character} and therefore the user's view into the world changes according to the physical \textit{VR Headset} movement. This is the ideal scenario for movement, where there is enough physical space available to the user to move anywhere she would want to go in the virtual world. However, this is hardly a realistic scenario since the models creatable in our solution can theoretically become arbitrarily large, meaning that the user would need an infinitely large physical space to move around in. Therefore, a secondary movement method, namely teleportation is required, that allows a user to move their \textit{Virtual Character} through the environment without moving in the physical world. Teleportation involves the user entering a teleportation mode, for example through pressing or holding a specific button on one of the \textit{VR Controllers}, and then aiming the controller at a spot that she wishes to teleport to. When either releasing the button or pressing it again the target position is selected and the \textit{Virtual Character} is instantly teleported to the aimed location. 
With these two movement methods---natural walking and teleportation movement---the player can reach any position in the \textit{Virtual Environment} independent of the size of physical space available to her while still moving in a rather natural way.

\subsection{\textit{3D Models}}

Usually, modeling languages are only specified with 2D visual representations. \ac{uml} is no exception from that rule. We could stick to those same 2D shapes inside a 3D world with users being able to position the 2D shapes freely in 3D. However, we believe using 3D shapes instead of 2D ones will likely result in a more natural experience for users because the real world only consists of 3D objects and we do not want the model elements to seem like foreign bodies in the \textit{Virtual Environment}. The main shapes used to visualize \ac{uml} elements in the 2D specification are rectangles and lines with different forms of arrowheads at the end of those lines. To ensure that users familiar with the 2D representations can learn the 3D ones easily we tried to find natural equivalents of the 2D shapes in 3D. The equivalent of rectangles in 3D are cubes and cuboids, while lines are best represented by thin tubes (see Fig. \ref{fig:user_avatar} (c)). A challenge of a 3D visualization of \ac{uml} Classes is the text representation inside them. A 2D Class only has one side that the text is displayed on which always faces the user. Since the Class rectangles known from 2D are most similar to Cuboids in 3D, every \ac{uml} Class is visualized as a cuboid in our solution (see Fig. \ref{fig:user_avatar} (a)). It can be assumed that the users can see at most three of the six sides of a cuboid at any given time. If the Class's cuboid only displays the text on one side, this side could, thus, be hidden depending on the view direction that the user has towards the Class. Rotating the cubes automatically so the text-side always faces the user would be a possible solution to that. However, we wanted the user to be the only entity changing the model's appearance, so the model seems more stable and thus natural. For these reasons, we chose to display the text associated with a Class on all sides of the Class. This means that every Class side basically shows the same 2D Class in a notation similar to 2D \ac{uml}. Since all these sides belong to a single Class it is important that all sides always show the same content.

\subsection{\textit{VR Text Input}}

In \ac{VR}, text input is more difficult compared to traditional computers due to the lack of a physical keyboard that provides haptic feedback on whether a key was hit or not. Since the user cannot see and use a real keyboard while in \ac{VR}, virtual keyboards are often used instead. There are two basic types of keyboards regularly found in \ac{VR} applications: Laser pointer and drum-style keyboards. Laser pointer style keyboards are usually displayed as a vertical plane in front of the user with each hand representing one laser pointer that can be used to press a key with a dedicated button. Since the buttons have to be rather large to be easily hittable, this input method requires a lot of space. An alternative to laser pointer keyboards are drum-style keyboards which we have chosen in our solution (see Fig. \ref{fig:user_avatar} (b)). These are shown in a horizontal, slightly tilted, form in front of the user like many real keyboards are as well. One virtual mallet is attached to each of the user's hands that can be used to hit a key similar to how a person would hit a drum in the real world. Because of the obvious drum analogy, it can be assumed that this control scheme is also easy to understand for most users and could be more natural to use than laser pointers.

\section{Implementation}

\begin{figure*}[ht!]
	\centering
	\includegraphics[width=\textwidth, keepaspectratio]{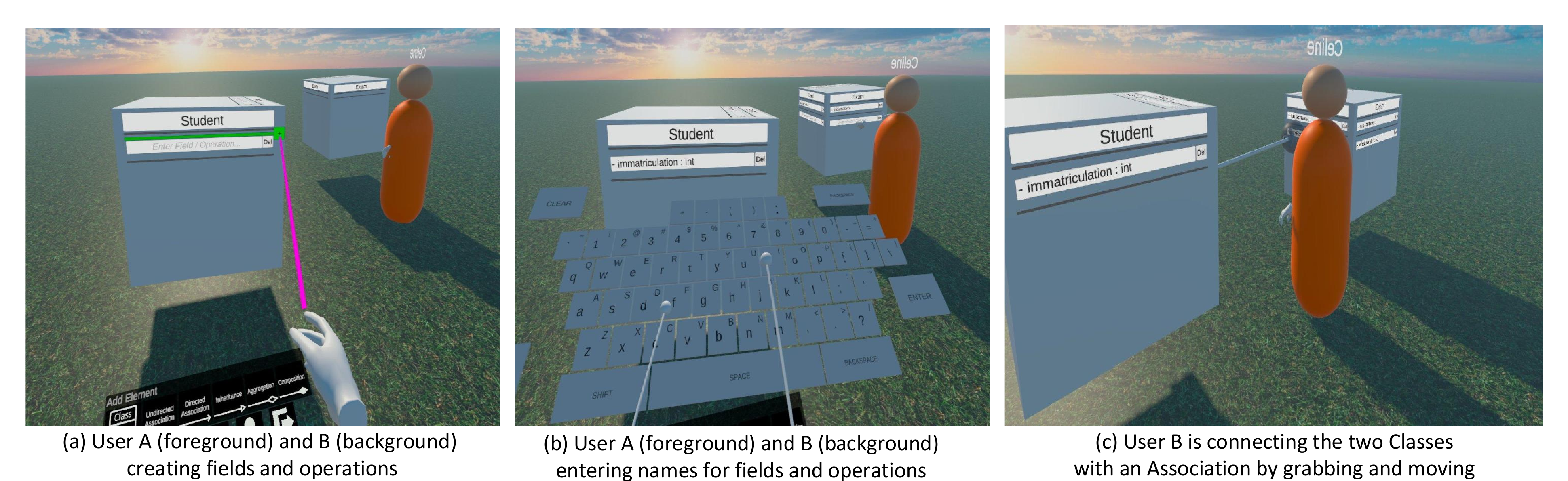}
	\caption{Screenshots of the collaborative VR modeling environment}
	\label{fig:user_avatar}
\end{figure*}

As a target implementation platform for our collaborative VR modeling environment, we have chosen the Oculus Quest 2 which is a cable-free, mobile VR HMD. For the implementation of the VR environment, we have used the Unity 3D Game Engine developed by Unity Technologies \cite{Unity3D} as it includes a fully featured Graphics, Sound and Physics Engine and has easy-to-use APIs for many different aspects like controller input. Furthermore, Unity also provides support for the Oculus Quest 2 among other Oculus VR headsets through an Oculus VR SDK. To enable remote collaboration, a networking system is needed that synchronizes aspects like model element- and user avatar positions, user hand gestures, etc. For the implementation of the networking system, we have used the \ac{pun} plug-in for Unity. This is a third-party system developed by Exit Games specifically for use in Unity multiplayer projects and offers easy-to-use high-level components. \ac{pun} realizes the network architecture discussed in the previous section by providing its own \textit{Cloud} servers that automatically work with \ac{pun} without the need for any custom server-side development. The Oculus \ac{sdk}, used to realize the interaction between the \textit{VR Device} and the \textit{Virtual Environment}, comes with a variety of assets that can be used to quickly implement common functionalities inside a \ac{VR} application. Most aspects needed to realize the \textit{Virtual Character} are covered by those assets provided by Oculus. Therefore, to save development time on those basics, we used Oculus' \textit{OVRAvatar} and \textit{OVRCameraRig} assets and adjusted them slightly for our application. The OVRCameraRig deals with tracking the position and rotation of the head and controllers and rendering the world accordingly into the headset. It does not include animated hands by default but there are separate assets for that which can simply be placed under the empty GameObjects that track the controller positions to enable users to see their hands. The gestures of the hands are tracked locally by default through an animator so we only had to add an animator synchronizer provided by PUN2 to synchronize the hand's animation states across the network \cite{PunPlayerNet}. Since all \textit{Virtual Characters} are represented equally in this prototype, we chose to display every user's name above their character. An example of how a remote user's \textit{Virtual Character} looks inside the \ac{VR} app is shown in Figure \ref{fig:user_avatar}.

\section{Evaluation}

To evaluate the efficiency, effectiveness, and user satisfaction of \textit{VmodlR} and to compare it with conventional collaborative modeling approaches, we have conducted a user study which will be presented in the following.

\subsection{Setup and Participants}

The user study was organized in such a way that two people in different rooms had to collaboratively create a UML Class Diagram. We have chosen a within-subjects design \cite{10.5555/2742543} for our user study where the participants were asked to use both modeling approaches, a conventional modeling tool, and the developed VR modeling tool. Two different small UML Class Diagram modeling tasks (consisting of five classes and based on a textual description) were provided to the participants, while the sequence and type of task which was carried with the help of a modeling tool were evenly distributed to avoid potential bias in the collected data.        
As a reference application for comparing with our own VR environment, we have decided to use an existing commercially available tool for collaborative \ac{uml} modeling. The tool we used is the Web application Lucidchart \cite{LucidCharts}. This was chosen because it offers a mode of collaboration many users are already familiar with from services like Google Docs, it offers a free version that could be used for this study, and supported \ac{uml} Class Diagrams. The evaluated applications focus on remote collaboration, therefore, we simulated a remote setting by placing each participant in a different room during their collaborative tasks where they could only communicate through computing devices (conventional and VR device). For the \ac{VR} application, each participant was positioned in a free space of approximately the same size and equipped with an Oculus Quest 2 \ac{VR} Headset including its two \ac{VR} Controllers. The participants were able to communicate over the application's  voice chat feature using the Oculus Quest's built-in microphone and speakers. The respective task was displayed inside the \ac{VR} app on a panel that could be opened and closed through an icon on the user's menu. The panel was positioned slightly to the left of the user so they could leave the panel open and edit the model in front of them at the same time. While using the Web task, each participant was sitting at a desk in the same room that the \ac{VR} free space was in, equipped with a laptop and a wired mouse. The participants could use the laptop's keyboard, its trackpad, and the mouse as they saw fit. Since the Web application does not include voice chat, the participants communicated via the voice chat application Skype that was running in the background on their laptops. The app used the built-in speakers and microphones of the laptops. The tasks were supplied to users on a sheet of paper so, like in the \ac{VR} app, it would not occlude their modeling environment unnecessarily. Since this study involved participants having to create two small Class Diagrams, these participants had to bring a basic knowledge about what a Class Diagram is and which purpose it serves for software modeling. Therefore, we had to rely on participants who either had lectures or school classes on this topic, for example in computer science lectures or subjects and/or who knew Class Diagrams from a different source, like working as a software developer who uses them. Therefore, we primarily tried to acquire participants with educational backgrounds, like university students and recent graduates. Our main source was a lecture on Model-Based Software Engineering designed for undergraduate students of computer science in their third year of studies. In this lecture, we presented the study and asked students to participate in it. We also reached out to students we were still in contact with who took part in this lecture during the prior year to widen our pool of possible participants while still ensuring comparable credentials among them. In total, 24 participants took part in the user study, meaning that there were 12 groups of two people each.

\subsection{Procedure}

The user study was conducted during one week in February 2021. The experimental setting was kept as equal as possible for all pairs of participants. First, the participants were greeted and introduced to the user study. Then, the basic procedure of how they will take part in the study and what they will do during their participation were explained. Afterward, depending on whether the \ac{VR} or Web app was used first, the collaboration environment was set up (e.g., splitting the participants, starting Skype, etc.), and they received a short introduction to the respective program. In the case of the Web application, this was done through the study supervisor explaining the main functionality from a pre-written script to ensure all participants were given the same information. In the \ac{VR} app, we have additionally implemented a tutorial that served as an on-boarding tutorial at the beginning of the user study. After the tutorials, the participants were provided with the respective task and instructed to solve the task collaboratively in the sense that they should only create one Class Diagram together in each application. They were then asked to indicate to the study executor once they think they are done with their task. When the participants finished the first task, the procedure was repeated for the second task. After both tasks were finished, the collaborative applications were closed so participants would not be able to talk to each other anymore. They were then given the questionnaire hosted via Google Forms \cite{GoogleForms} and were asked to fill it out using the laptops that were used for the Web application as well. During this process, the users stayed in the different rooms they were in while working on the tasks to ensure that they would not influence each other's answers.

\subsection{Usability Measurements}

\begin{figure}[ht!]
	\centering
	\includegraphics[width=0.4\textwidth, keepaspectratio]{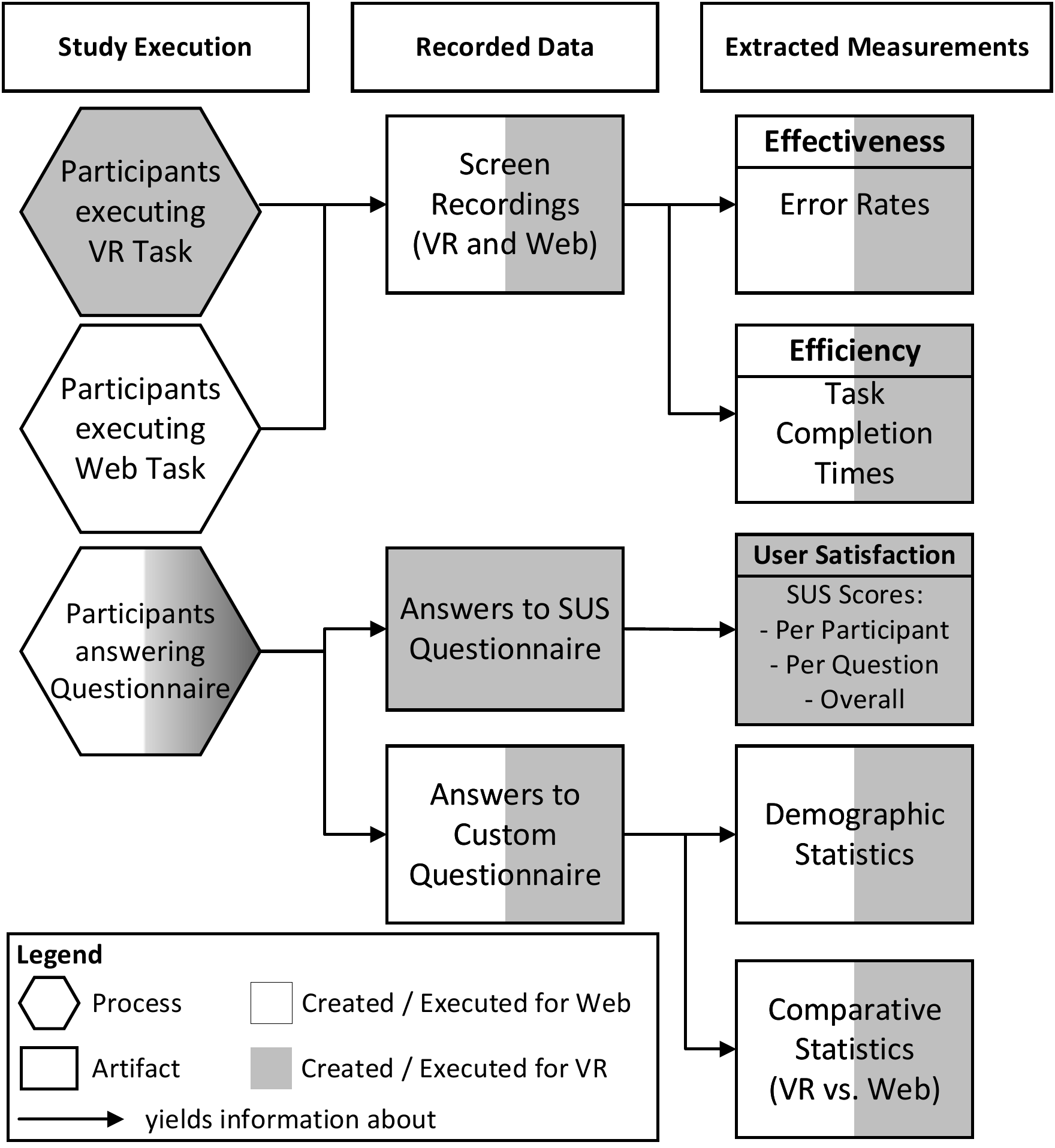}
	\caption{An overview of what was measured within the user study.}
	\label{fig:measurements}
\end{figure}

To extract meaningful results from the study we had to choose certain measurements that we would take during the execution. Figure \ref{fig:measurements} shows an overview of the data that was collected during the user study and which measurements were derived from that.

Efficiency was measured by recording the participants' execution of the task and tracking the time from the point where they started reading the task to the time when they told the study executor that they finished it. From these same recordings, the effectiveness was measured by counting the number of operating errors that the participants made during the execution of the task. From this, an error rate was calculated by dividing the number of errors by the time in minutes that was measured as the efficiency. This error rate was our final score for the effectiveness in errors per minute. The recordings used for tracking efficiency and effectiveness were screen-captures including the voice chat audio. In each group, only one participant was recorded to reduce the amount of video data that had to be manually evaluated. Another reason for that was the network infrastructure that did not allow us to capture and record two video feeds from the \ac{VR} headsets in parallel. The user satisfaction was evaluated through a questionnaire that participants were asked to fill out after they finished both tasks. We chose to use the \ac{sus} questionnaire \cite{brooke1996} since it is a well-proven and reliable questionnaire that provides comparability with \ac{sus} evaluations of other applications and that can be quickly filled out due to its low number of questions \cite{Bangor2009}.
While efficiency and effectiveness were measured for both applications, the \ac{sus} questionnaire was only asked in the context of the \ac{VR} application. This is because the \ac{sus} questionnaire's main purpose in this work is to provide a proof of concept that the \ac{VR} solution has a rather good usability. Furthermore, the participants were asked to answer a custom-developed questionnaire that included mainly 5 point Likert scale questions (like the \ac{sus} questions) and some free text questions asking for more specific impressions with regard to collaboration and interaction. 

\subsection{Results}

In the following, we present the main results of the user study.

\subsubsection{Demographic Statistics}

24 people in 12 collaborating groups participated in the study. All participants were between 22 and 30 with a median of 25 and a mean of 25.5. 21 of the 24 participants were male and the others female. Most of the participants were software engineering practitioners or had a background in computer science. As a consequence, many participants reported advanced experience with UML and UML modeling tools. The average rating scores for prior experience with UML and UML modeling tools were 3.25 and  approximately 3, respectively. With regard to prior experience with VR, our results show that half of the participants have never used a VR device before this study while 8 used them at least reasonably often.  

\subsubsection{Efficiency}

\begin{figure*}[ht!]
	\centering
	\includegraphics[width=\textwidth, keepaspectratio]{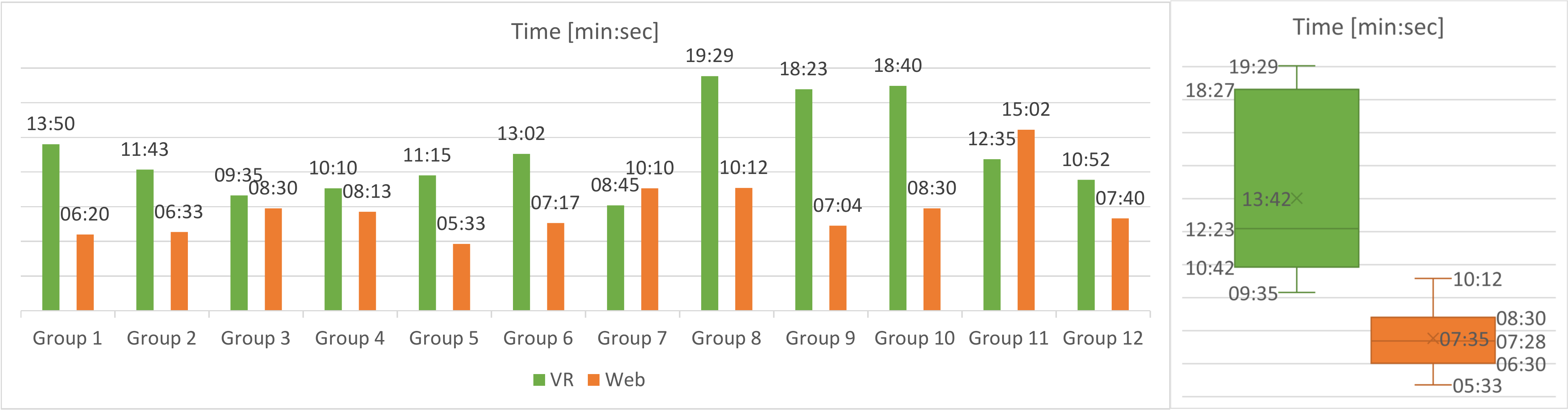}
	\caption{The times each group took to execute the task in the Web and the \ac{VR} app respectively.}
	\label{fig:TimeDiag}
\end{figure*}

The times participants needed to complete their tasks are shown in Figure \ref{fig:TimeDiag}.

All but two groups (Group 7 and 11) needed more time in \ac{VR} than in the Web application to complete their task. From the recordings, it could be observed that almost all groups split up the work on the Class Diagram by each modeling one part of the diagram and putting both parts together in the end. During this process, they occasionally discussed how to model certain aspects if they were unsure and checked the part of the model their collaborator created in the end. The two groups that took longer in the Web application than in \ac{VR} had a slightly different approach that could explain this anomaly: In \ac{VR} they split up their task as outlined above. In the Web application, however, they mainly followed a pair-programming style approach: For each part of the task they discussed how to model it, and then only one of them modeled the aspect accordingly. The first approach obviously saves time in comparison to the latter one which could explain why they were able to complete the task faster in \ac{VR} than in the Web application when everyone else needed longer in \ac{VR}. This makes the times for these two groups not comparable since they did not use a similar organizational method in both conditions. Therefore, we excluded them from time-related analyses like the box plots shown on the right of figure \ref{fig:TimeDiag}. Besides this anomaly, it can be said that the spread with respect to time was way smaller for the Web application than for \ac{VR}, i.e. they vary less as shown on the right of figure \ref{fig:TimeDiag}. In \ac{VR}, times range from almost 20 minutes to as low as 9:30 min, while in the Web condition, the times vary from approximately 5:30 to 10 minutes. The difference of each group between Web and \ac{VR} had a mean of 4:46min and a median of 5:26. This time-data may be normally distributed around its mean but does not have to be. To perform a significance test, we used the Wilcoxon Signed-Rank test \cite{Arif2017} since it does not assume a normal distribution and based on our test we cannot confirm that both samples are normally distributed. This test reports a \(Z = -2,589\) with \(p < .05\), meaning that \ac{VR} had statistically significantly longer task times than Web.

\subsubsection{Effectiveness}

In order to consistently track errors across all recordings we had to define what "error" means in the context of this usability evaluation. Our definition of a \textit{Usability Problem} is based on \cite{Manakhov2016} where the authors define a \textit{Usability Problem} as "a set of negative phenomena, such as user's inability to reach his/her goal, inefficient interaction and/or user’s dissatisfaction, caused by a combination of user interface design factors and factors of usage context" \cite{Manakhov2016}. Based on this definition, we extracted 3 types of errors that we analyzed in the recordings:

\begin{enumerate}
	\item Missed Interaction Point: \textit{The user tried to interact with an element of the application but did not hit said element (for example a button) or used the wrong control for interaction. Thus, the user needs to repeat the interaction.}
	\item Accidental Interaction: \textit{The user did not intend to interact at all or not with this specific element but accidentally interacted with it anyway. Thus, the user needs to revert the interaction.}
	\item False Interaction: \textit{The user tried to do an interaction that is not possible at all or not possible at that specific element. Thus, the user experiences a loss of time and needs to find out the correct interaction to achieve the desired effect.}
\end{enumerate}

\begin{figure*}[ht!]
	\centering
	\includegraphics[width=\textwidth, keepaspectratio]{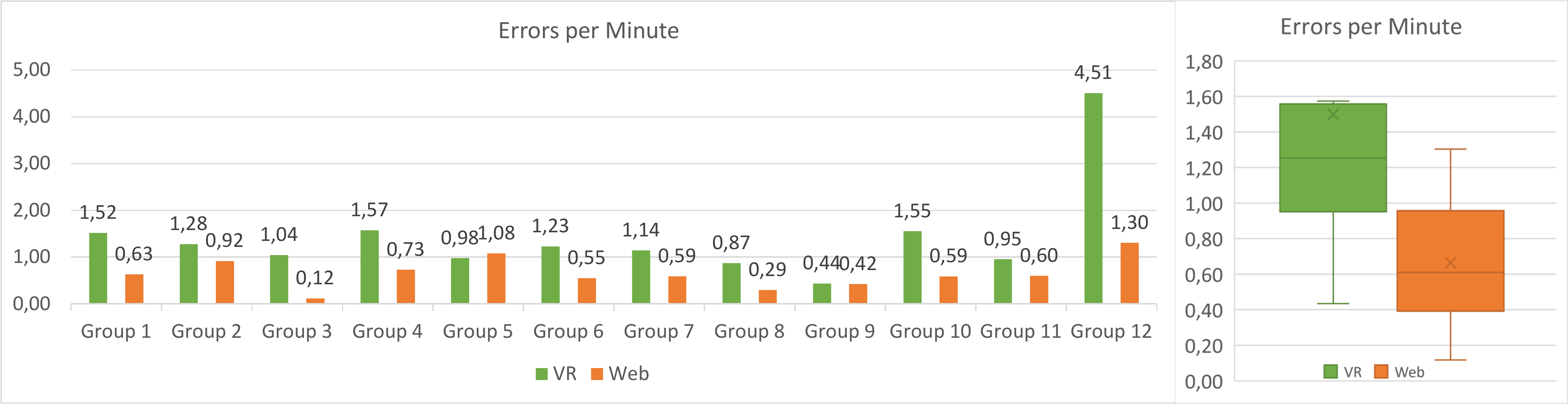}
	\caption{The error rates of the recorded participant in each group.}
	\label{fig:ErrorRates}
\end{figure*}

During the evaluation of the recordings, we tracked each error by its type and a time stamp. Finally, the effectiveness was measured in errors per minute and is shown in Figure \ref{fig:ErrorRates}. Most errors belonged to the \textit{Missed Interaction Point} or \textit{Accidental Interaction} types with only a few \textit{False Interactions}.

In \ac{VR}, we observed a mean error rate of approximately 1.5 (error per minute) with a median of 1.25 (error per minute), while these values lied at 0.66 and 0.61 (error per minute) respectively in the Web application. 
The Wilcoxon Signed-Rank test resulted in a test statistic of \(Z = -2.599\) with a significance \(< .05\) which shows that \ac{VR} had a statistically significantly higher error rate than Web.

\subsubsection{User Satisfaction}

The basic user satisfaction was measured in the \ac{VR} application using the 10 items \ac{sus} questionnaire for each participant. In total, our collaborative VR modeling environment reached an average SUS score of 78 out of 100 which indicates according to \cite{Sauro2011} good usability. As mentioned earlier, we did not ask the participants to fill out the \ac{sus} questionnaire for the Web application as our focus was more on the acceptance of our own VR solution than on the acceptance of a mature and commercial modeling tool like Lucidchart. 

\subsubsection{Additional Questionnaire Results}

\begin{figure}[h]
	\centering
	\includegraphics[width=0.44\textwidth]{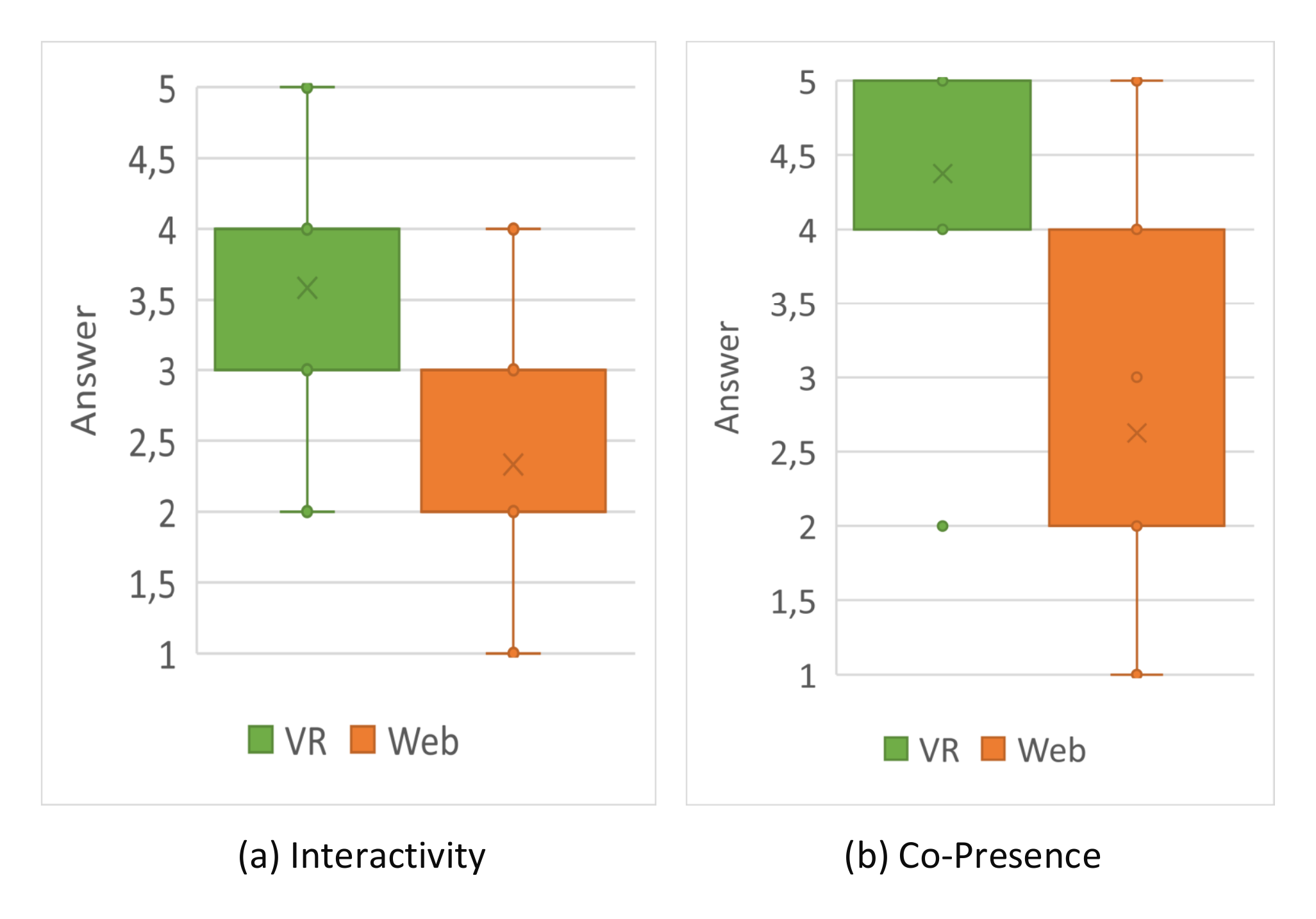}
	\caption{Results of additional questions concerning interactivity and co-presence}
	\label{fig:FaceToFaceSimilarity}
\end{figure}

To assess the perceived naturalness of both applications, the Web and our VR solution, the similarity of interactions inside the applications to face-to-face interactions and the feeling of co-presence were analyzed. The results are depicted in Fig. \ref{fig:FaceToFaceSimilarity}. 

Concerning the similarity of interactions to face-to-face interactions (see Fig. \ref{fig:FaceToFaceSimilarity} (a)), the average score of the Web application is 2.33 and noticeably lower than the \ac{VR}'s 3.58. Additionally, in \ac{VR} 14 out of 24 people selected the highest (5/5 on the Likert scale) answers whereas, in the Web, nobody chose the highest and only two people chose the second highest answer (4/5 on the Likert scale). A Wilcoxon Signed-Rank test revealed that this observed difference is indeed statistically significant with \(Z = -3.208\) and a significance of \(< 0.05\). Overall, this means that users found interactions with their teammates to be significantly more similar to face-to-face interactions in the \ac{VR} modeling environment compared to the Web application. A further important aspect for natural interaction and collaboration is co-presence which denotes the feeling of being in the same place despite the remote setting. Fig. \ref{fig:FaceToFaceSimilarity} (b) shows how the participants assessed their feeling of co-presence in both applications. It shows that in the Web, this feeling is very different from user to user, with larger bulks at both ends of the scale. The median, however, with a value of 2.6 points more towards the lower end of the scale. In the \ac{VR} app, this is very different. While two people rated the co-presence rather low (2 on the 1-5 scale), every other participant at least somewhat agreed with the sentiment that there was a feeling of co-presence with 9 participants selecting 4/5 and 13 choosing 5/5. The Wilcoxon Signed-Rank test confirmed that this difference is statistically significant with a \(Z = -3.587\) and a significance of \(< 0.05\). This implies that the \ac{VR} application can more successfully make users feel like they are collaborating in the same room compared to the Web app.  Apart from the above-mentioned questions, we asked the participants to provide us general feedback and remarks (what have you liked/disliked most) on the developed collaborative VR modeling environment. The most notable result is that 14 out of 24 people mentioned that moving the model elements through the grabbing feature felt natural. Furthermore, most comments were praising the "collaboration" in the VR modeling environment. Some comments explicitly mentioned a like for the collaboration (3 out of 24), some indicated a positive impression for the ability of talking to the teammate (4 out of 24), being able to see the teammate (4 out of 24) or the feeling of the teammate's presence (4 out of 24). Some participants mentioned multiple of these aspects. In total, 11 out of 24 noted that at least one of these collaboration-related aspects felt natural or intuitive. Concerning the negative feedback comments, typing on the keyboard in VR was the most mentioned aspect (9 out of 24) while some of the participants stated that this would only need some time to get used to it. Furthermore, the missing of an auto-alignment or snapping feature, like known from most diagramming and modeling tools (e.g. Microsoft Visio \cite{MSVisio}), was complained about by five users. However, five out of 24 people said that \ac{VR} was entertaining or fun to use and two added that it is specifically useful for home office scenarios since it makes people feel more together and would "definitely improve motivation and team spirit". Finally, the participants had to choose which application they would prefer to use for collaborative \ac{uml} modelling: They could choose either one or state that they would want to use both depending on the situation. A follow up question to this one subsequently asked in which situations they would want to use which application if they selected "Both". 13\% (3 out of 24) wanted to use the \ac{VR} app over the Web application and 37\% (9 out of 24) vice versa. Half of the participants, however, indicated that they would want to use both applications, each for specific situations. When asked in which situations they would want to use which application, participants gave rather diverse answers. Five users stated that they would want to use the Web application for complex or longer tasks and the \ac{VR} one for shorter ones. Two mentioned that they would prefer the Web application for time critical work. Brainstorming and planning was mentioned by three people to be more suited for the \ac{VR} application and one person stated that she would like to use "the web application when you are working on one device with your partner" and the \ac{VR} app when in "different locations".

\subsubsection{Discussion}

Based on previous research and the participants' familiarity with traditional computer programs, we expected one downside of \ac{VR} to be that users' task executions are slower and more error-prone in \ac{VR}. The data from our study shows that this was indeed the case. These measures could have been influenced by the universal familiarity of participants with PC applications in general and \ac{uml} tools specifically. It is therefore possible that speed and error rates in \ac{VR} improve as users get more experience with a specific \ac{VR} \ac{uml} tool. However, the current state-of-art text entry can be seen as a major bottleneck for the efficiency of text-intensive VR applications. The results of the \ac{sus} evaluation additionally showed that our \ac{VR} implementation is already quite usable even though it still lacks many features that users expect from such an application like automatic aligning of model elements and copy \& paste functionalities. The various data points gathered about the naturalness of different aspects of the application gave a clearer insight on what concrete advantages such a \ac{VR} application can have compared to traditional tools: On average, users found the interactions and especially the collaboration related aspects of the \ac{VR} application significantly more natural than in the Web application. This is especially important with respect to this work's focus on remote collaboration settings as the study showed that the feelings of being together and collaborating face to face with a co-worker were much higher in \ac{VR} compared to the traditional PC alternative. Another aspect that we expected \ac{VR} to be beneficial for is the motivation and fun users are having while using it. Our study shows that it is indeed true that users were a lot more motivated and had a lot more fun using the \ac{VR} application compared to the Web app. It is important to note that this could be influenced by the fact that \ac{VR} is a relatively new and therefore possibly more interesting technology, so these values might align more over time when a user regularly uses a \ac{VR} application for modeling. The fact that using \ac{VR} can be more exhausting for people was one downside of \ac{VR} we expected to observe in this study as well. While we did see that some participants experienced significant discomfort (due to heavy headsets, pressure against the face, or cybersickness), most people did not have any issues with that. Considering the suggestions made by participants about when they would like to use which application, it is evident, however, that a \ac{VR} \ac{uml} tool would not simply replace the desktop ones. Rather it would be an amendment, so users have the option to use the tool that is most suited for a given situation. In remote collaboration settings, where two or more people need to brainstorm or discuss how something should be modeled, the \ac{VR} application could be used. When collaborating in the same room or when creating a model alone, a PC-based application could be more appropriate. This implies that such apps need seamless interoperability between PCs and \ac{VR} when they ought to be used in actual modeling work outside of usability studies. Summarizing, it can be said that \ac{VR} can offer a more natural collaborative modeling experience compared to PC-based tools but that both techniques have certain advantages and drawbacks. These make the use of both tool-types, depending on the concrete situation, most sensible instead of using only one of them exclusively.

\subsubsection{Threats to Validity}

With regard to collaboration, this study only evaluated two-person teams. Therefore, it is unclear if the findings can be generalized to larger numbers of simultaneous collaborators. The number of participants and thereby collaboration teams was also rather limited. With more participants, a between-subjects design \cite{10.5555/2742543} could have been chosen which has the possibility to provide data and thus findings that are more generalizable across many different users. Due to the participant's demographics (e.g., age, profession, background, etc.), the findings of this study should be taken carefully and further user studies with other user groups are needed to get insights about the general usability. Additionally, this study compared a fully-featured commercially available Web application with a \ac{VR} modeling prototype, missing many of the features that the Web application supported, like automatic aligning of model elements. This study therefore could not achieve a strict like-for-like comparison between the two types of applications.

\section{Conclusion and Future Work}

While the Unified Modeling Language (UML) is one of the major conceptual modeling languages for software engineers, more and more concerns arise from the modeling quality of UML and its tool-support. Among them, the limitation of the two-dimensional presentation of its notations and lack of natural collaborative modeling tools are reported to be significant. In this paper, we have explored the potential of using Virtual Reality (VR) technology for collaborative UML software design by comparing it with classical collaborative software design using conventional devices (Desktop PC / Laptop). For this purpose, we have presented a VR modeling environment that offers a natural collaborative modeling experience for UML Class Diagrams. Based on a user study with 24 participants, we have compared collaborative VR modeling with conventional modeling with regard to efficiency, effectiveness, and user satisfaction.

In future work, we plan to extend our collaborative VR modeling environment to further UML modeling diagram types and even to further modeling languages (e.g., BPMN, SysML, etc.). In addition, further improvements of the VR modeling environment are planned to support a multi-modal (e.g. speech in-/output or haptic feedback) interaction and a more realistic representation of the virtual character (e.g. realistic avatars or tracking of users’ facial expressions). Features like automatic aligning of model elements, and tracking of which collaborator created which parts of the model are also useful aspects that could be integrated into the VR modeling environment. In addition, further usability evaluation studies with larger groups of heterogeneous participants and more complex modeling tasks should be conducted to analyze in more detail the benefit of collaborative modeling in VR. Finally, we believe that a cross-device mixed reality collaborative modeling approach is a promising way to support modeling across different AR/VR capable and conventional devices.   

\bibliographystyle{./bibliography/IEEEtran}
\bibliography{./bibliography/IEEEabrv,./bibliography/IEEEexample,./bibliography/Paper}
\newpage

\end{document}